# Machine and Component Residual Life Estimation through the Application of Neural Networks

M.A. Herzog[a], T. Marwala[b] & P.S. Heyns[a]

[a]Dynamic Systems Group, University of Pretoria, Pretoria, Republic of South Africa
[b]Control and Systems Group, University of Witwatersrand, Johannesburg, Republic of South Africa

This paper concerns the use of neural networks for predicting the residual life of machines and components. In addition, the advantage of using condition-monitoring data to enhance the predictive capability of these neural networks was also investigated. A number of neural network variations were trained and tested with the data of two different reliability-related datasets. The first dataset represents the renewal case where the failed unit is repaired and restored to a good-as-new condition. Data was collected in the laboratory by subjecting a series of similar test pieces to fatigue loading with a hydraulic actuator. The average prediction error of the various neural networks being compared varied from 431 to 841 seconds on this dataset, where test pieces had a characteristic life of 8,971 seconds. The second dataset was collected from a group of pumps used to circulate a water and magnetite solution within a plant. The data therefore originated from a repaired system affected by reliability degradation. When optimized, the multi-layer perceptron neural networks trained with the Levenberg-Marquardt algorithm and the general regression neural network produced a sum-of-squares error within 11.1% of each other. The potential for using neural networks for residual life prediction and the advantage of incorporating condition-based data into the model were proven for both examples.

**Key Words:** Neural Networks, Condition Monitoring Data, Residual Life

## 1 Introduction

The advent of preventive maintenance has increased the need for reliable information, leading to the development of data analysis techniques for the purpose of estimating residual life. The traditional approach, described by Coetzee [1], involved the use of probabilistic models which were fitted to data on failure times but more recently, researchers such as Pijnenburg [2] have investigated the use of regression models which allow explanatory variables to be incorporated. Condition-based data is commonly available and Vlok [3,4] found that its use enhanced the accuracy of the predictions made by regression models. Accurate residual life estimates have a number of benefits for tactical maintenance planning, apart from the selection of an optimal replacement strategy and the flexibility offered to the maintenance manager. Such information allows the advanced planning of shut-downs, resource allocation and the optimal holding of spares.

A different approach is required for renewal and repaired systems, which are not returned to a good-as-new condition after failure. In the renewal case, it is assumed that, once repaired, the system is returned to its original state. If a system is not repaired to its original condition, this assumption does not hold and system deterioration due to imperfect repair has to be taken into account. Reinertsen [5] states that a considerable number of papers have explored the estimation of residual life for renewal systems through the use of statistical methods, but no corresponding work has been done on repaired systems that do not conform to the assumption that they have been returned to their original state. Pijnenburg [2] comments on the extreme rarity, in the literature he reviewed, of datasets on repaired systems, in which failure times are listed in the original chronological order. Ascher and Feingold [6] could find only four such datasets.

The research studies using condition-monitoring data for residual life estimation include the work of Jantunen [7] who fitted a polynomial curve to vibration data, Vlok [3,4] who used regression curves and vibration data to estimate the residual life of pumps, and Wang and Zang [8] who used spectrographic oil analysis data to predict the residual life of aircraft engines. Vlok [3,4] found that regression models offered a significant advantage over parametric models because of their ability to take into account the information relating to the failure of a system. Condition-monitoring data could therefore be used to improve the accuracy of the estimates made with these models.

Research has been done to investigate the use of neural networks in applications related to maintenance and reliability. A wide variety of methods, network architectures and data combinations were used in these cases. Amjady and Ehsan [9] evaluated the reliability of power systems using an expert system based on neural networks. Luxhøj and Shyur [10] compared the performance of traditional reliability modeling techniques with neural networks for the fitting of a reliability curve to the data of helicopter components. Luxhøj [11] researched the prospect of providing FAA safety inspectors with a means to evaluate and control the appropriate surveillance levels for aircraft operators through the use, among other things, of neural networks. Liang, Xu and Shun [12] applied MLP neural networks to the field of condition monitoring, whereas Xu et al. [13] attempted to forecast reliability by using neural network techniques to analyze the data on past historical failures. Neural networks have therefore been employed for maintenance-related applications, but their use has not yet been fully explored in the context of residual life prediction. As these networks have the capacity to learn about the underlying relationship between various inputs and outputs, they are ideally suited to making predictions about complex systems.

This research builds upon the work done by others who employed regression models for predicting failure. As an alternative to traditional statistical methods, this study investigates the suitability of neural networks for making reliability predictions in the cases of both renewal and repair. The incorporation of covariates containing historical information and condition data into the training process is explored with the aim of improving the accuracy of the predictions that can be made. The performance of different neural network types when trained with reliability data is also of interest, and the results achieved by a selected group of networks are compared. Based on these results, conclusions can be drawn on the suitability of using neural networks in conjunction with condition-monitoring data for reliability predictions as part of the tactical planning done by the maintenance practitioner.

## 2 Problem description

### 2.1 Renewal dataset

The first dataset represents the renewal case where the system is returned to a good-as-new condition by replacing the failed component. A series of laboratory tests were conducted, using a 630kN Schenck Hydropuls hydraulic actuator (see Figure 1), which simulated an actual maintenance situation encountered in industry. A number of similar notched test pieces were manufactured with the same cross-sectional shape as a component which serves as an overload protection in jaw crushers. This toggle plate is designed to fail when foreign objects become wedged between the crusher jaws, thereby preventing damage to the machine. The test pieces were placed under a cyclic loading in the hydraulic test rig until they failed as a result of fatigue. The cyclic loading was applied according to a sinusoidal pattern, where the mean and amplitude were varied by means of the actuator's control system, in this way generating different operating conditions for the series of test runs and producing a varied dataset. Though the actuator is capable of exerting a maximum force of 630 kN, the actual applied load pattern was selected to ensure that the components had a finite fatigue life.

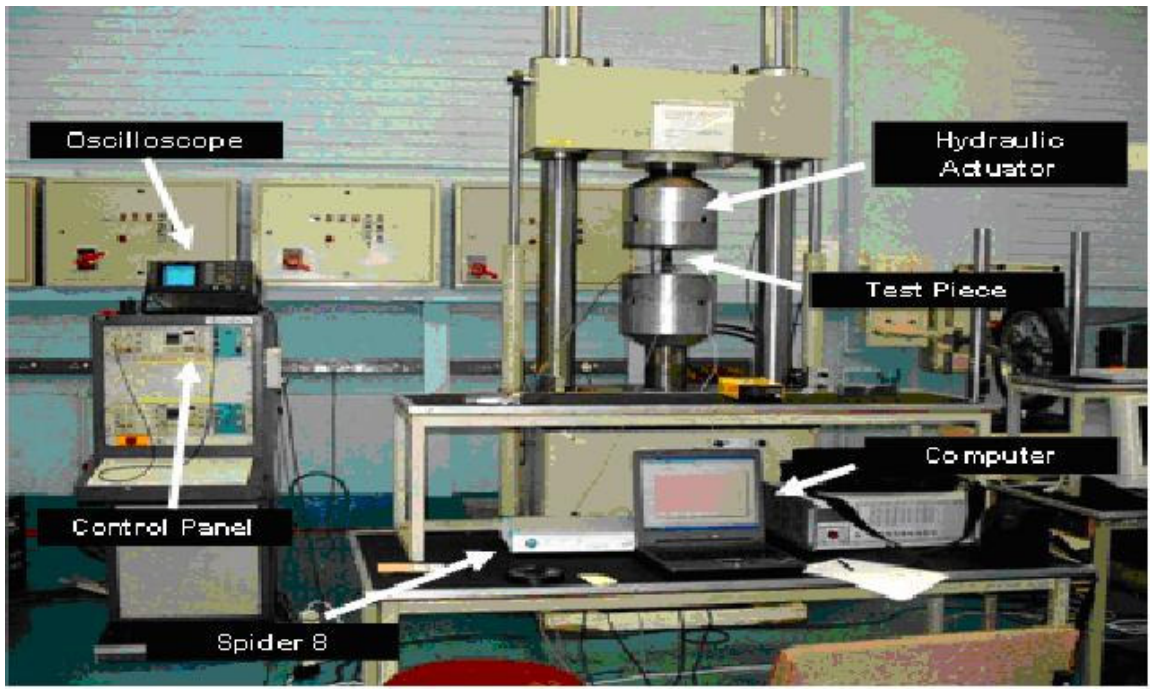

**Figure 1: Photograph showing the various elements involved in the lab testing.**

The actuator of the test rig was set to maintain constant amplitude in the oscillation of its jaws for the duration of each test run. The amplitude was varied for the different test runs, thus altering the operating conditions to which each test piece was subjected. A specific initial load could be applied by increasing the displacement of the jaws at the start of a test run until the required load cell reading was attained. As the cracking of the test piece in the notch area caused weakness, the force required to maintain the amplitude was reduced and this could be observed in the corresponding drop in the magnitude of the load cell measurements that were taken. The reduction in applied force resulting from the use of displacement control provides a measurable indication of deterioration in the condition of the test piece.

The loading pattern was applied at a frequency of 3Hz which was close to the upper limit of what could be achieved while still allowing the actuator to apply a suitably high load. Measurements were recorded over periods of three seconds at three-minute intervals. The testing proved that the time interval between the taking of measurements and the duration of the recording window were both satisfactory. At a frequency of 3 Hz, the data for a total of nine complete actuator cycles was captured in each measurement window, in which a sequence of 1,800 samples was taken during the three-second period.

Four different sensors were selected and used for taking the measurements during each such measurement window. The choice of sensors was not only aimed at tracking the deterioration in the test piece deterioration, but also at providing a measure of the operating conditions that influenced the life of the test piece.

Measurements were taken with the load cell forming part of the test equipment, as well as with a strain gauge attached to the test piece. These sensors provided information on the nature of the applied load. An accelerometer was mounted on the opposite side of the location of the notch. The purpose of this measurement was to measure the movement due to the deflection of the test piece under loading.

The temperature on the surface of the test piece was measured by means of a thermocouple mounted on the side of the test piece. For convenience, the thermocouple was positioned halfway along the cross-section and aligned with the center of the notch. It was found that the temperature measured at this position rose dramatically once crack propagation started. The temperature measurement was therefore found to be a useful indicator of test piece condition and gave a good indication of imminent failure. The magnitude of the rise in temperature compared with the initial measured temperatures was dependent on the applied loads and therefore also served as an indicator of the rapidity with which failure was occurring.

### *2.2 Repaired system dataset*

As an example of a repaired system, a dataset was used which had been obtained by Vlok [3,4] from the Sasol Twistdraai mine plant. Measurements were taken on eight identical Warman pumps used to circulate a water and magnetite solution within the plant. Four main failure modes were identified for these particular pumps, namely bearing seizure, broken or defective impellers, damaged or severely eroded pump housings, and broken drive shafts. The measurements taken on the pumps were solely vibration readings, for which a spectral analysis was performed and a number of fault frequency bands were monitored. The frequency bands 0.4×RPM, 1×RPM, 2×RPM, and 5×RPM were monitored for both the bearings of these pumps. Measurements were unfortunately only taken sporadically and the dataset is therefore sparse.

During the 791-day window from the initial installation of the eight pumps, pump operation was suspended eight times due to condition-based warnings, and 11 failures were recorded. The surprisingly high percentage of failures might be attributed to the inconsistent application of the condition-based policy and the long measurement intervals. Although the data was collected from the start of each pump's life, the vibration measurements were taken extremely infrequently. Vlok [3,4] does note that some of the failures occurred suddenly, with deterioration occurring in a matter of hours. Obviously, it would be difficult to predict such a sudden deterioration with the information that was available. From the random nature of the measurements, it appears that the final measurement ahead of the suspension of a pump's operation may have been prompted by clearly observable external signs of pump deterioration.

Three of the eight pumps experienced only one failure, two of the pump units failed twice, and three units each failed four times. On average the pumps lasted 469 days to the first failure or preventive intervention. This can be compared with an average of 134 days, 103 days and 137 days to the second, third and fourth failures or preventive interventions, respectively. Reliability therefore deteriorated dramatically after the first failure, indicating imperfect repair. A further pattern was observed with regard to the time to the first failure, and this pattern allowed the pumps to be subdivided into two groups. Pumps which failed for the first time after more than 500 days, tended to fail only once or twice during the period in question. The remaining units averaged 357 days to first failure and each failed four times within the time window.

# 3  Neural network application

The MATLAB neural network toolbox was used to build and train the neural networks for the purpose of residual life prediction. A number of network variations in terms of architecture and training algorithms are available in this programming environment.

## 3.1  Network testing

The usefulness of a neural network in a practical application depends on the degree to which it can generalize when confronted with data which was not seen during training. Methods have been developed to test and compare the performance of different networks with this aim in mind. Schenker and Agarwal [14] identify the three most common methods for testing the relative performance of neural networks:

- A subdivision of the available data into a training and test set, termed a static split.

- Cross-validation, which can be described as a dynamic split of the data.

- Statistical evaluation without splitting the data.

Testing through the use of statistical methods, according to Schenker and Agarwal [14], is only meaningful when the data represents a true process. It can therefore be successfully applied in cases involving reliable physically based models . Schenker and Agarwal [14] identify the subdivision of the dataset into separate training and test sets as the approach that is most commonly used, even though only part of the dataset can be used for training which limits this method's application to larger datasets. In their comparison of the performance of the different testing methods, Schenker and Agarwal [14], in their comparison of the performance of the different testing methods, point out that a strategy of cross-validation generally outperformed such a static split in the search for an optimal network for a particular application.

For the purposes of comparing different neural network variations by cross-validation, the dataset is broken into a number of smaller groups which do not overlap. These groups are cyclically allocated to the training and the test sets. Each cycle in the cross-validation process represents a completely independent training run, so that the networks are not tested with data used for training at a previous stage. The error on the test data is recorded for each of the network variations at the completion of each cycle. Several partially overlapping portions of the available data are therefore used for training the neural networks, but each group of data is used only once for testing. The recorded error values are added once the process has been completed, and this result is the basis for comparing the different neural networks.

The greatest advantage of using cross-validation is that the entire dataset can eventually be used for training the neural network once the optimal neural network layout has been found. The loss of information due to a static split of data is therefore avoided, which is important in cases where the dataset is limited in size. Training does unfortunately become more cost-intensive due to the repetition required for cross-validation.

### *3.2 Neural network for renewal dataset*

The first-order gradient descent learning algorithm serves here as the basis for comparing the different neural networks due to its historical significance. Adjustments were made to the learning rate, and a momentum term was introduced that increased the rate of convergence of this algorithm. The performance of the gradient descent algorithm was compared with the much faster second-order Levenberg-Marquardt algorithm which, according to the findings of Hagan and Menhaj [15], outperformed other fast techniques. Bayesian regularization (see Bishop [16]) was applied in conjunction with the Levenberg-Marquardt algorithm to investigate the effect of this method which is aimed at improving generalization. The general regression neural network (GRNN), which was also used by Luxhøj [11] in his research, has the advantage of rapid unsupervised training. It is also of interest because it is a network with radial basis function (RBF) architecture, in contrast to the MLP architecture of the networks mentioned so far.

A static split was chosen as the method for comparing network performance on the renewal dataset. This was feasible because of the simplicity of the simulated maintenance setup in the laboratory, for which there was only one failure mode. The lab data collected during testing was split into two groups: nine of the datasets were used for training and the remaining three comprised the testing set.

Each network was constructed with five inputs and generated a single output which represented an estimate of remaining life to failure, measured in seconds. The MLP networks were each constructed with five nodes in the hidden layer, so that the basic network structure was similar for each of these networks. The size of the hidden layer was optimized through an empirical process where the number of nodes in the hidden layer was varied.

The inputs used for network training were the elapsed time of the specific test at the time of the measurement, initial average load, initial load range, change in load range, and change in temperature. The network inputs and outputs were normalized and transformed into values between zero and one.

Elapsed time gives the network an indication of the component's age and allows the network to differentiate between new samples, and samples that already show fatigue. Therefore the network can differentiate between two samples which are subjected to the same loading but do not yet exhibit measurable signs of deterioration.

The network is given a longer-term predictive capability by providing it with information about the operating conditions to which the test piece is subjected. The initial load average and range define the conditions to which the test sample was subjected during testing. The network is therefore trained to differentiate between test pieces subjected to higher and lower loading, which is the main factor contributing to the rapidity with which failure occurs.

The changes from initial load and temperature give an indication of deterioration in the condition of the test piece and impending failure. Due to the setting of the machine, displacement remained constant and therefore the load dropped when cracking started. Temperature increased substantially as fatigue damage worsened and the crack propagated through the test piece. Therefore the network can make adjustments to its prediction once overt signs of impending failure become apparent. This adjustability allows the network to cope more easily with unexpected events and changing conditions.

### *3.3 Neural network for repaired system*

As the sparseness of the pump dataset did not allow for the use of a separate test set, it was decided that cross-validation should be used to test the performance of different network designs. To this end, the dataset was divided into eight groups, each representing the data from one of the pumps. In their work, Schenker and Agarwal [14] assert that individual runs should not be split when using cross-validation, as this would violate the assumption that the test and training sets are independent. The total life of each pump was therefore deemed to be one run and the data was grouped accordingly.

On the basis of the performance of the neural networks that were trained with the renewal dataset, it was decided that the focus should be on the network types that could be trained more rapidly, as cross-validation involves the time-consuming repetition of network training. Accordingly, the standard Levenberg-Marquardt algorithm, the Levenberg-Marquardt algorithm with Bayesian regularization, and the GRNN were chosen for comparison.

The actual data was pre-processed in a similar way to the renewal dataset. It was found that the high values measured at an advanced stage of deterioration led to a distortion in the normalized data inputs used to train the neural networks. The neural networks became insensitive to the small changes occurring in the initial stages of deterioration. As the aim of this work is not to prove the usefulness of condition-based maintenance, but to improve longer-term predictions of expected life, the readings taken during the last week before the occurrence of failure were discarded. This decision led to an improvement in the accuracy of predictions at earlier stages of deterioration.

The use of a greater number of network inputs representing condition-based information is expected to improve the network's ability to make accurate predictions. To test this hypothesis, each of the neural network layouts was trained with three, four and five inputs. The first set of training runs was done by using the elapsed time since installation, the elapsed time since the last failure, and a covariate that can be described as a risk variable dependent on the history of the pump. Two further training runs were completed, first with one and then with two additional inputs, each of which represented the average value of the vibration response amplitude in a chosen frequency band for the measurements on the two bearings. Using the findings of Vlok [3,4] as basis, additional inputs based on condition related measurements should improve the accuracy of failure predictions.

The dataset originates from a repaired system and its reliability is therefore affected by previous failures and repair. The influence of these factors has to be taken into account, even though not much of this information was recorded. Vlok [3,4] states that alarm levels were used as prescribed by the pump manufacturer, but these values are not given and the cause of failure or the reason for a condition-based suspension and overhaul was not indicated. An empirical risk variable was consequently based on the observed pattern which indicates that pumps that required an early repair tended to fail more frequently. For the data collected before the occurrence of the first failure, the risk variable $R$ is set equal to 1. After the first failure, Equation 1 is used to calculate the value of $R$.

$$R = \tfrac{1}{2}\left(\frac{T_1}{T}\right)^2 \tag{1}$$

The risk variable $R$ is therefore reduced to 0.5 immediately after the first failure and its value decreases at a rate dependent on $T_1$ which is the time to the first failure. $T$ is the elapsed time since the initial installation of the pump unit. A large value of *R* therefore corresponds to a low risk of failure, whereas a small value indicates a high risk. It takes into account the significant reduction in reliability after the first failure and the characteristic of a high failure rate in cases where an early first failure is recorded.

The hidden nodes of the MLP networks were varied according to the number of inputs presented to the networks to test the effect of such changes in network structure on network performance. Training was firstly done for networks with the same number of inputs and hidden nodes. Then a second training run was done with a hidden layer that had one node more than the input layer. Due to ill-conditioning, however, MLP networks with six hidden nodes could not be trained with five inputs. The dataset size used for cross-validation contained 53 data points. Once this had been subdivided into groups, the largest group contained 13 data points, which meant that the smallest training set would contain 40 data points. The maximum number of hidden nodes in an MLP network with five inputs was therefore limited to five, in order to prevent ill-conditioning as discussed by McKeown et al. [17], because the number of variables in the network exceeded the number of inputs. The networks all generated a single output, namely a prediction of the remaining life until the next failure of the pump.

## 4  Neural network results

### 4.1  Neural network results for renewal data

The traditional way of conducting a data analysis on the reliability data originating from a renewal system is to fit a statistical distribution to such data. This technique, described by Coetzee [1], was accordingly chosen to form the basis of comparison to illustrate the advantage of using neural networks.

A Weibull distribution was fitted to the data of the training set and the parameters of the two-parameter Weibull distribution were found to be $\beta = 1.7522$ and $\eta = 8971$. The Weibull parameter η is the scale parameter, which is also referred to as the characteristic life. Coetzee [1] notes that 63.2% of components fail before this time and 36.8% survive. The use of a statistical distribution means that no specific prediction can be made about an individual test piece. The estimated life is therefore taken as the characteristic life of the whole population of the training set. The actual residual life for the test sets differed by between 11.2% and 55.4% from the characteristic life of 8,971 seconds, that was calculated using this statistical method. The results achieved by fitting the Weibull distribution show the disadvantages of this method when comparing them with the residual life results obtained by using neural networks, which are discussed in the following paragraphs. Table 1 shows accuracy of the life predictions on the test set, using the different methods with the data available at the start of the various test runs.

**Table 1: Accuracy of predictions for the test data with initial measurements recorded at the start of the experiments.**

| Network | Test data |
|---|---|
| Weibull | 11.2% – 55.4% |
| GDBP with M | 4.0% - 34.9% |
| LM | 1.9% - 20% |
| LM with BR | 3.1% - 5.1% |
| GRNN | 1.6% - 68.9% |

The standard back-propagation algorithm was used to train the same network architecture with nine different combinations of the learning rate ($\alpha$), and momentum parameter ($\beta$). Figure 2 illustrates the rate of convergence of the gradient descent back-propagation algorithm with a different combination of training parameters. Oscillations become much more pronounced when a higher learning rate is used and training clearly becomes much more rapid. If the learning rate is increased even more, the training process becomes unstable, overshoots the target and no convergence on a minimum is achieved. The training process must therefore balance the rate of convergence with the requirement of stability.

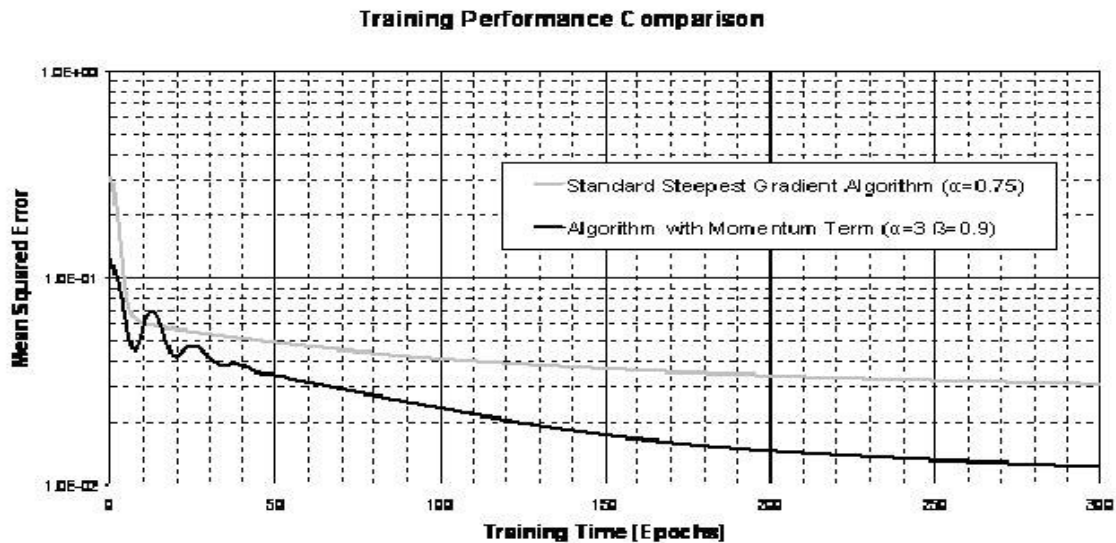

**Figure 2: Comparison of the rate of convergence of the gradient descent back-propagation method, using a different combination of training parameters.**

It was found that a learning rate of 0.75 and a momentum constant of 0.9 gave good results, so these constants were used for the comparison with other network types and training algorithms. The training algorithm was stopped early and could not accommodate some of the more isolated data points in the training set. It was therefore possible to maintain improved properties of generalization.

Training with the Levenberg-Marquardt algorithm proved much more rapid and a far better fit was achieved after less than 300 training epochs. The neural network's estimated residual life for the training data was within 5% of the actual remaining life of each component, when presented with the first recorded inputs after the start of the test run. The largest error was 449 seconds, which compares very favorably with the 180-second interval between measurements, which is the band within which the failure occurred. The network performance on the training data is therefore a satisfactory result. The accuracy of the prediction obtained by the neural network on data that had not previously been seen during training, was similar in two of the cases. The largest error on the test set was 20%, which indicates some degree of overfit, as the data from this particular test piece had was least similar to the training data than the three test sets did.

It was expected that the use of Bayesian regularization would address the problem with overfitting encountered with the network trained with a standard Levenberg-Marquardt algorithm. The neural network that was trained with Bayesian regularization did not produce as close a fit for some parts of the training set as the fit achieved with the standard Levenberg-Marquardt algorithm. In particular, the estimates generated for some of the isolated data points on the training set displayed a large error. This was expected, as the regularization technique penalized training in order to maintain the network's capability of providing acceptable results for new data. The two test pieces in question had a significantly shorter life than the other test pieces and were therefore isolated from the rest of the data. The benefit of this regularization technique regarding improved generalization becomes clear when examining the results obtained for the test set. The largest error in a network prediction for the data of the test set was 5.1 %. The prediction of the neural network in this case was only 513 seconds adrift of the actual recorded life of 10,102 seconds. The results achieved for all three of the test pieces in the test set were therefore very satisfactory and indicated a good generalization.

Setting up the GRNN is an almost instantaneous process, due to its adaptation of the input vectors for the hidden nodes, and the use of the target vectors as weights in the output layer. No supervised training is required in its construction. Network performance can therefore be influenced only by changing the value of the bias of the radial basis function nodes in the hidden layer.

The bias of a radial basis function in MATLAB is set by defining a parameter, called the spread value. Every bias in the first layer of the network is set to 0.8326 divided by this spread value. The radial basis functions in these neurons therefore have an output of 0.5 when the absolute value of the distance between the input and weight vectors is equal to the spread. The area of the input space to which each neuron responds is thereby set where the spread alters the radius of the basis functions, and therefore determines the amount of overlap and consequently the smoothness of the fit.

When designing an RBF network, it must be ensured that the spread of the RBF neurons is large enough. If the radial basis function neurons overlap enough, several radial basis function neurons will generate significant outputs at any time. The resulting network function is smoother and a better generalization is achieved for new input vectors that fall between the input vectors used in the design of the network. If the overlap is too large, however, too many neurons will then react to every input, and accuracy will be forfeited. A number of different spread values were tested, and the results are tabulated in Table 2.

Table 2: Performance of the GRNN for different spread values.

| Spread | MSE training | MSE test |
|---|---|---|
| 0.01 | $9.9 \times 10^{-7}$ | 0.0030 |
| 0.02 | $3.8 \times 10^{-6}$ | 0.0027 |
| 0.03 | $8.3 \times 10^{-5}$ | 0.0022 |
| 0.04 | $8.5 \times 10^{-4}$ | 0.0026 |
| 0.05 | $2.3 \times 10^{-3}$ | 0.0034 |

Figure 3 illustrates that the larger the spread chosen for the network, the smoother the fit. The quality of the fit on the training set is reduced, as a number of hidden layer neurons start to influence the output for any given input. But an increase in spread improves network generalization until an optimal balance is reached. Any further increase in spread is detrimental to network performance.

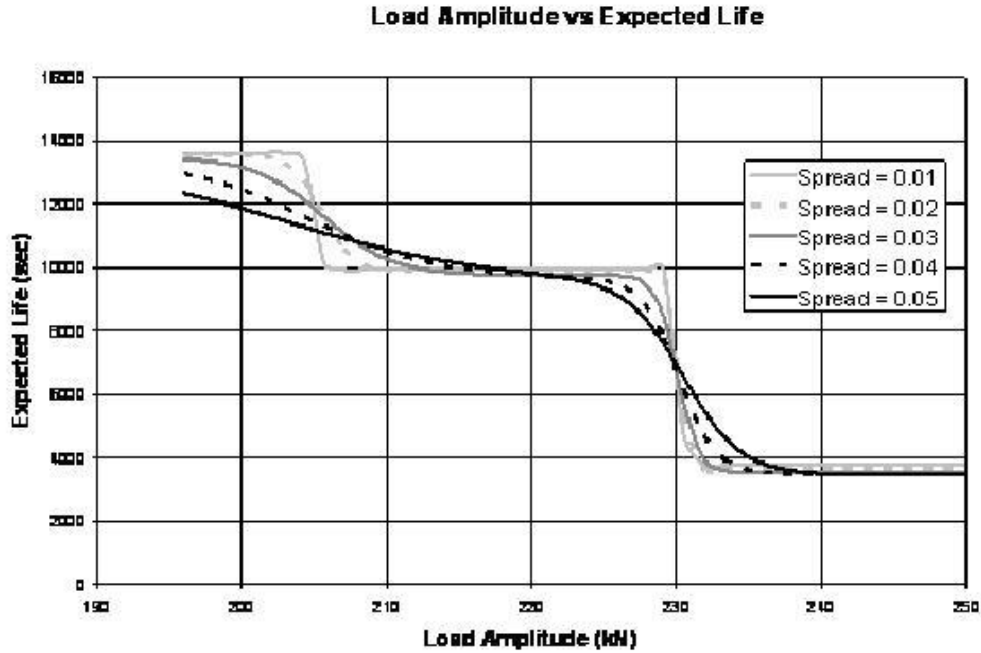

**Figure 3: GRNN response when varying the load amplitude input.**

The estimated residual life for the training data when using a small spread value was closer to the actual life than was achieved with any of the other networks, when using the MLP architecture and supervised training. This can be attributed to the way in which the GRNN is trained and the insignificant overlapping of nodes with a small radius. An exact fit is expected in this particular case, as the network should respond with the expected target vector if provided with a training vector. As was the case with the standard Levenberg-Marquardt algorithm, overfitting occurred during the design of the GRNN and a large error in the residual life estimate was observed for one of the test pieces.

**Table 3: Comparison of the mean squared error (MSE) on the training and test sets of the different networks.**

| Network | MSE training | MSE test |
|---|---|---|
| LM with BR | $5.7 \times 10^{-5}$ | 0.0014 |
| GRNN | $8.3 \times 10^{-5}$ | 0.0022 |
| LM | $8.1 \times 10^{-5}$ | 0.0030 |
| GDBP with M | $1.9 \times 10^{-2}$ | 0.0061 |

When comparing the performance of the different networks, it was found that the MLP network trained with a Levenberg-Marquardt algorithm using Bayesian regularizations had a clear advantage over the other models. The gradient descent algorithm was found to be significantly slower than the Levenberg-Marquardt algorithm. The advantage of an unsupervised training process, which was mentioned in the literature, was proven by the speed with which the GRNN could be trained. Network learning in this case proved instantaneous.

Table 4 shows the average prediction error whereas Table 5 gives the maximum prediction error of the networks being compared. Though the GRNN has a lower average error than the other networks, it has the highest maximum error. This may explain why the MLP network trained with the Levenberg-Marquardt algorithm with Bayesian regularization outperformed it in terms of the mean squared error. The early stopping of the gradient descent back-propagation algorithm meant that higher maximum and average errors were recorded for the training set. This phenomenon can be ascribed to the sparseness of the dataset, which led to isolated data in the problem space. As the training algorithm was stopped before it could accommodate this data, the network performed well on the test data.

**Table 4: Comparison of the average error in the predictions made by the networks.**

| Network | Training data | Test data |
| --- | --- | --- |
| LM with BR | 64 sec. | 455 sec. |
| GRNN | 87 sec. | 431 sec. |
| LM | 84 sec. | 616 sec. |
| GDBP with M | 1370 sec. | 841 sec. |

When considering these results, it should be borne in mind that the measurements were taken at intervals of 180 seconds, and that the time of first measurement after failure was used as failure time for training the neural networks. The actual failure took place within the band spanning the last measurement cycle. All the neural networks performed very well on the data in the test set that was closest to the training data. The data for the test piece, showing the greatest variation from anything the network had seen before, proved to be the greatest test of each network's ability to generalize. The advantage of using Bayesian regularization to improve the network's ability to generalize is clearly illustrated when comparing the graphs of the results relating to this series of data.

**Table 5: Largest error in the predictions made by the networks being compared.**

| Network | Training data | Test data |
|---|---|---|
| LM with BR | 513 sec. | 1065 sec. |
| GRNN | 411 sec. | 3085 sec. |
| LM | 652 sec. | 2185 sec. |
| GDBP with M | 3997 sec. | 1271 sec. |

The results for the GRNN are not as smooth as those obtained with MLP networks. The jagged shape of the prediction graphs illustrates the "local" nature of RBF networks, compared with the "global" nature of MLP networks. This property may adversely affect network performance, if the information for one set of the data points used for training, are corrupt. A greater overlap of the RBF nodes would counteract this situation by smoothing the transition between kernels.

The results prove that neural networks can be successfully employed to make reliability predictions for a renewal system. When presented with the first set of measurements collected after the start of a test run, all the neural networks generated predictions which were more accurate than the results obtained through the traditional statistical method of fitting a Weibull distribution to the failure data. In particular, the accuracy of the predictions made by the MLP trained with the Levenberg-Marquardt algorithm with Bayesian regularization would be suitable for making maintenance decisions in the context of the simulated situation.

## *4.2 Results for the Repaired System*

The neural networks trained with the Levenberg-Marquardt algorithm used nodes with the log-sigmoid transfer function in both the hidden and output layers. During initial training with a small mean-squared-error training target of $1 \times 10^{-5}$, it was found that overfitting occurred and the neural networks failed to generalize the test data. A series of training runs with a range of different training targets were consequently performed in order to improve generalization by terminating the training process at an earlier stage. The training targets used for this purpose were the values $1 \times 10^{-5}$, $5 \times 10^{-3}$, $1 \times 10^{-2}$, $2.5 \times 10^{-2}$ and $5 \times 10^{-2}$.

Table 6 shows that the best results obtained with the smallest error on the test data, were achieved with the larger target values $5 \times 10^{-2}$ and $2.5 \times 10^{-2}$. The comparatively small error on the training data indicates that the training algorithm was stopped before the overfitting characterizing the worst results (shown in Table 7) could occur. The target values refer to the normalized output values, whereas the sum-of-squares error is calculated from an error value in days.

**Table 6: The best results achieved with the Levenberg-Marquardt training algorithm.**

| Network architecture | Inputs | Target | $\Sigma$ (error)$^2$ |
|---|---|---|---|
| 5 hidden nodes | 5 | 0.05 | $2.70 \times 10^5$ |
| 5 hidden nodes | 4 | 0.025 | $2.85 \times 10^5$ |
| 4 hidden nodes | 4 | 0.025 | $2.90 \times 10^5$ |
| 3 hidden nodes | 3 | 0.05 | $2.92 \times 10^5$ |
| 4 hidden nodes | 4 | 0.05 | $2.92 \times 10^5$ |

**Table 7: The network results with the largest error after training with the Levenberg-Marquardt algorithm.**

| Network architecture | Inputs | Target | $\Sigma$ (error)$^2$ |
|---|---|---|---|
| 3 hidden nodes | 3 | 0.00001 | $4.32 \times 10^5$ |
| 4 hidden nodes | 3 | 0.00001 | $4.45 \times 10^5$ |
| 4 hidden nodes | 3 | 0.01 | $4.49 \times 10^5$ |
| 5 hidden nodes | 5 | 0.01 | $4.99 \times 10^5$ |
| 5 hidden nodes | 5 | 0.00001 | $5.31 \times 10^5$ |

As the network error usually did not reach the smaller target values of $1 \times 10^{-2}$, $5 \times 10^{-3}$ and $1 \times 10^{-5}$, the training process was terminated once the pre-set limit of 100 epochs had been reached. These networks consequently suffered from overfitting and failed to perform well on the test data.

Changing the size of the hidden layer and the number of inputs was affected by the early stopping of the training process, so that no clear pattern emerged. Though an additional node in the hidden layer was beneficial when training towards an error target of $2.5\times10^{-2}$, it seemed to be detrimental when training with a target value of $5\times10^{-2}$. It did appear that a greater number of inputs generally improved the performance of these networks, but the results were not conclusive.

The MLP neural networks trained with the Levenberg-Marquardt algorithm with Bayesian regularization (LMBR) yielded similar results to the networks trained for the optimal duration with the standard Levenberg-Marquardt algorithm. In this case the Bayesian regularization prevented overfitting during training, thereby improving the network's ability to generalize.

The log-sigmoid transfer function was used for the nodes in the hidden layer of these networks, whereas two different transfer functions were utilized in the output layer. Table 8 summarizes the results achieved with the neural networks trained with the Levenberg-Marquardt algorithm with Bayesian regularization.

**Table 8: Levenberg-Marquardt algorithm with Bayesian regularization.**

| Network architecture | Inputs | $\Sigma$ (error)$^2$ |
|---|---|---|
| 5 hidden nodes, linear output node | 5 | $2.81\times10^5$ |
| 5 hidden nodes, sigmoid output node | 5 | $2.90\times10^5$ |
| 5 hidden nodes, linear output node | 4 | $2.95\times10^5$ |
| 4 hidden nodes, sigmoid output node | 4 | $3.06\times10^5$ |
| 5 hidden nodes, sigmoid output node | 4 | $3.07\times10^5$ |
| 4 hidden nodes, linear output node | 4 | $3.15\times10^5$ |
| 4 hidden nodes, linear output node | 3 | $3.26\times10^5$ |
| 3 hidden nodes, linear output node | 3 | $3.35\times10^5$ |

| Network architecture | Inputs | Σ (error)² |
|---|---|---|
| 3 hidden nodes, sigmoid output node | 3 | $3.52 \times 10^5$ |
| 4 hidden nodes, sigmoid output node | 3 | $3.52 \times 10^5$ |

The results shown in Table 8 indicate that the choice of transfer function of the output node had a far greater influence on the performance of the network than the variation in the number of nodes in the hidden layer. Another observation is that in this case, additional input information clearly leads to more accurate predictions.

The GRNN networks were trained with RBF neurons with different sensitivities so as to select an optimal value for this parameter. Table 9 lists the results for the networks with values of 0.1 and 0.05 for the spread parameter.

In contrast to the other network types, the best results with GRNN were achieved with four inputs but an increase in input space to five inputs led to overfitting. The consequence of an increase in the number of inputs into such a network is that the outputs of the nodes are influenced by a greater number of variables, hence making them more sensitive to a particular combination of input values. Once this sensitivity becomes too great, the network starts to lose its ability to generalize. For this reason, the general regression neural networks did not respond well when presented with five inputs.

**Table 9: Cross-validation for GRNN**

| Network architecture | Inputs | Σ (error)² |
|---|---|---|
| Spread = 0.1 | 4 | $3.00 \times 10^5$ |
| Spread = 0.05 | 4 | $3.32 \times 10^5$ |
| Spread = 0.1 | 3 | $3.56 \times 10^5$ |
| Spread = 0.05 | 3 | $3.72 \times 10^5$ |
| Spread = 0.1 | 5 | $3.72 \times 10^5$ |
| Spread = 0.05 | 5 | $4.17 \times 10^5$ |

Two different values were used for the spread in the GRNNs. It was found that the less sensitive networks with a spread of 0.1 generally achieved better results. The decreased sensitivity achieved with a larger radius for the basis function led to a reduced degree of overfitting in the network.

The ease of implementation of the GRNN was again illustrated. The construction of this type of network is instantaneous, as no weight adjustments are made by implementing a back-propagation algorithm. By varying the sensitivity of the RBF neurons, adjustments can be made to optimize the network's ability to generalize. An optimal network can therefore be rapidly found by employing cross-validation.

In summary, when testing network generalization by means of cross-validation, the best results obtained with the various neural networks were very similar, once these networks had been optimized in respect of this particular dataset. Table 10 gives a comparison of the best results achieved with each network type.

**Table 10: Comparison of the best results achieved by the different network types.**

| Network architecture | $\Sigma$ (error)$^2$ |
|---|---|
| LM | $2.70 \times 10^5$ |
| LMBR | $2.81 \times 10^5$ |
| GRNN | $3.00 \times 10^5$ |

The comparison of the different networks by cross-validation was based solely on the relative size of the sum-of-squares error obtained on the test data. If the suitability of the applied method should be judged, the results should also be viewed in the context of the practical application. It was found that number of very large prediction errors were made by the networks on isolated points, which far exceeded the actual remaining time to failure of a specific pump. When the ten worst predictions were excluded, the average prediction error of the networks was 39.8% for the LMBR network, 33.2% for the GRNN and 41.7% for the LM network.

The nature of this result indicates that the networks were able to model some but not all of the significant properties of these complex pump systems. When seen in the context of the intended application, the results represent a positive point of departure. An average prediction error of 40% is too large and does not allow these networks to be used in their current form for making decisions about maintenance. It can therefore be said that the complexity of the problem requires a larger and more descriptive dataset for training the neural networks, if more accurate results are to be obtained.

A key element in the successful practical application of neural networks is to find suitable covariates which will allow the network to distinguish among different scenarios and failure modes. The smallness of the dataset also has the result that part of the data in the test set will in some cases differ significantly from the data with which the network was trained. The network is therefore unable to deal with some of the data correctly, and produces a spurious result. The dataset used by Vlok [3,4] is not ideal for this purpose owing to its sparseness, and it is probable that the given data could not achieve much more regarding failure prediction.

Despite these deficiencies, it was proved that it is possible to combine the advantages of failure time data analysis and condition monitoring in a neural network platform to make more accurate predictions.

One should bear in mind the limitations imposed on residual life predictions by the unpredictability of operations in an actual plant. The covariates chosen as inputs into a neural network have to reflect the failure modes of the system. If a failure cannot be traced by one of these inputs, it will be impossible for the network to predict more accurately when the machine will fail. The results achieved in this study can therefore be seen as a conditional success in terms of the use of neural networks for this application.

# 5 Conclusion

The use of neural networks for making failure predictions for both renewal and repaired cases was investigated. The estimates that the networks made regarding the simulated renewal system proved highly accurate, with the average error varying between 431 seconds and 841 seconds for the different types of neural networks. This compares well with the measurement interval of 180 seconds which was used. It was shown that much greater accuracy could be achieved with neural networks than with the use of the common probabilistic technique that involves fitting a Weibull distribution to the failure-time data. The performance of the neural networks was compared with this statistical method on the basis of the predictions made when the networks were presented with the first set of values, measured on the test pieces allocated to the test set. The MLP neural network trained with the Levenberg-Marquardt algorithm using Bayesian regularization did not exceed a prediction error of 5.1%. By comparison, the error of the residual life estimates made using the Weibull distribution, ranged between 11.2% and 55.4%.

The failure predictions for the repaired systems were hampered by the combination of the system's complexity and the sparseness of the dataset, however. The sparseness of the dataset limits the number of inputs that can be used for MLP networks and also means that the input space is poorly mapped. Repaired systems have multiple life intervals that are not independent and are subject to numerous failure modes, posing a severe challenge to the analyst. The small number of inputs and poorly mapped input space meant that the explanatory information proved insufficient for the network to model the system accurately, and large errors were recorded on some of the test data.

With respect to the comparison between different neural network methods, the use of Bayesian regularization proved very effective in the prevention of overfitting. The use of a second-order method, such as the Levenberg-Marquardt training algorithm, produced a significant reduction in training time in comparison to the gradient descent method. It was found that the optimization of network parameters was an important part of the training process and that the performance of different network types was very closely matched once their design had been adjusted to suit a specific application. GRNN are simple, easily generated neural networks and proved a close match with the MLP networks, giving a difference of 11.1% on the sum-of-squares error for the repaired system dataset.

The ease with which neural networks can be trained and the quality of the results achieved for the two datasets indicate that neural networks should become a useful tool for the analysis of reliability data in future. Clearly the approach outlined in this paper is not suitable for every application in the maintenance field, but the results indicate the potential that neural networks have as a powerful tool for the analysis of reliability data and the prediction of residual life.